\documentclass{article}
\usepackage{spconf,amsmath,graphicx}
\usepackage{booktabs}
\usepackage[htt]{hyphenat} 
\usepackage{cite}
\usepackage{xcolor}
\usepackage{pgfplots}
\usepgfplotslibrary{fillbetween}

\title{Using Speech Synthesis to Train End-to-End Spoken Language Understanding Models}

\name{Loren Lugosch$^{1,3}$, Brett Meyer$^{1}$, Derek Nowrouzezahrai$^{1,3}$, Mirco Ravanelli$^{2,3}$}
\address{$^1$McGill University, $^2$Universit\'e de Montr\'eal, $^3$Mila}
\begin{document}
%
\maketitle
\begin{abstract}
End-to-end models are an attractive new approach to spoken language understanding (SLU) in which the meaning of an utterance is inferred directly from the raw audio without employing the standard pipeline composed of a separately trained speech recognizer and natural language understanding module. 
The downside of end-to-end SLU is that in-domain speech data must be recorded to train the model. In this paper, we propose a strategy for overcoming this requirement in which speech synthesis is used to generate a large synthetic training dataset from several artificial speakers.  Experiments on two open-source SLU datasets confirm the effectiveness of our approach, both as a sole source of training data and as a form of data augmentation. 

\end{abstract}
\section{Introduction}
\label{sec:intro}
The use of end-to-end models for spoken language understanding (SLU) is beginning to be given more serious consideration \cite{tomashenko2019recent, Haghani2018, bhosale2019end, caubriere2019curriculum}.
Whereas conventional SLU uses an automatic speech recognition (ASR) component to transcribe the audio into text and a natural language understanding (NLU) component to map the text to semantics, an end-to-end model maps the audio directly to the semantics \cite{Qian2017, Serdyuk2018, Chen2018}. End-to-end models have several advantages over the conventional SLU setup: they have reduced computational requirements and software implementation complexity, avoid downstream errors due to incorrect transcripts, can have the entire set of model parameters optimized for the ultimate performance criterion (semantic accuracy) as opposed to a surrogate criterion (word error rate), and can take advantage of information present in the speech signal but not in the transcript, such as prosody.

But because the input to an end-to-end model is speech and not text, end-to-end models cannot learn from text data. This means that new audio data
must be recorded to train the model for every new SLU domain or application. In contrast, the conventional ASR-NLU pipeline can be trained just once on a generic speech corpus to learn the mapping from speech to text, and subsequently only on text data. Thus, end-to-end SLU can be more difficult to implement in practice than conventional SLU because audio is more expensive and time-consuming to obtain than text data.

In this paper, we propose a method for reducing, or avoiding entirely, the need to record audio data to train an end-to-end SLU model. Given a dataset of semantically labeled text data, we use a generic speech synthesizer, or text-to-speech (TTS), to read out these texts, thus generating an audio dataset that can be used for training the model. The ability to use synthetic data greatly lowers the barrier to entry for people who want to develop an SLU model for a new application: even if the accuracy of a model trained on synthetic speech is not satisfactory for end users, it may be good enough to allow fast prototyping of voice interfaces without waiting on the slow, expensive process of recording real speakers. 
Our method is useful not only when no real data is available: it also acts as data augmentation by exposing the model to more speaking styles and more ways of pronouncing the same phrases.

Our main contributions in this paper are as follows:\footnote{The PyTorch code for our experiments is available online at \texttt{https://github.com/lorenlugosch/end-to-end-SLU}.}
\begin{itemize}
    \item We show that it is possible to train an end-to-end SLU model using only synthetic speech and achieve high accuracy on a test set of real speech, even when the speech synthesizer has imperfections.
    \item We run experiments using synthetic speech to augment an existing dataset of real speech and show that this augmentation can significantly improve accuracy, especially when few real speakers are available.
\end{itemize}

\section{Related Work}
\label{sec:related_work}

Our method is closely related to the idea of using speech synthesis to generate training data for end-to-end ASR \cite{li2018training, rosenberg2019speech}. In end-to-end ASR, instead of using a separate acoustic model, language model, and pronunciation model, a single sequence-to-sequence model predicts the transcript from the audio \cite{graves2014towards}. Because the language model in end-to-end ASR is only implicit and not decoupled from the rest of the model, it is difficult to train on standalone text data, so it does not easily handle certain types of utterances that are not well represented in the training audio, such as numeric sequences \cite{Peyser2019,He2019}. To help the model recognize these domain-specific types of utterances, they can be synthesized and added to the training set.

Outside of speech recognition, backtranslation is another technique in a similar vein often used for data augmentation in machine translation \cite{poncelas2018investigating}. In backtranslation, given three languages $A$, $B$, and $C$ and paired data for $(A,B)$ and $(B,C)$, synthetic paired data for $(A,C)$ is generated by translating the $B$ text in $(B,C)$ data into language $A$ using a model trained on $(A,B)$ data, and vice versa. If we think of the three modalities of audio, text, and semantics as three ``languages'', then our proposed technique is just backtranslation from semantically labeled text into audio.

Another related idea is ``sim2real'' transfer in robotics \cite{jakobi1995noise}. In sim2real transfer, a policy is learned in a simulated environment, avoiding the risks involved in physically operating a robot, such as breaking the robot or harming humans in the environment. The speed of simulation can also give the robot more experiences than would be possible in a limited amount of time in the real world. Likewise, fast speech synthesis can allow generating more audio than would be possible with a human speaker, due to time constraints or fatigue for the speaker.

\section{Proposed Method}
\label{sec:proposed_method}
The method proposed in this paper is simple. Two ingredients are required: 1) a text dataset, where each example consists of a transcript (e.g., ``turn it up a couple notches'') and corresponding semantic label (e.g., \texttt{\{"intent": "ChangeVolume", "slots": [\{"action": "increase"\}, \{"amount": "two"\}]\}}), and 2) a TTS for the language in which the transcripts are written. The TTS is used to synthesize each transcript. The label assigned to the synthesized audio is the label for the transcript used to synthesize the audio. If the TTS has multiple speakers, each speaker is used to synthesize the transcript, so that multiple training examples per transcript are generated. A subset of the available speakers can be used for a given transcript if it is too expensive to use all speakers. If spoken training examples from real speakers are available, the real and synthetic datasets can be concatenated to form a single larger dataset. An end-to-end SLU model can then be trained using the generated dataset.

We have identified three criteria that are important for choosing the TTS: 
\begin{enumerate}
     \item \textit{Multi-speaker:} In the past, we have found that having multiple speakers in the training set is crucial to achieving high test accuracy in end-to-end SLU. We anticpated that this would also be the case when using synthetic speakers.
    \item \textit{``Everyday'' voices:} Commercial TTS voices typically speak in refined ``actor speech'', which is pleasant for the listener. But this type of speech sounds very different from the casual speech in which most people naturally speak to voice interfaces. To avoid this mismatch, casual, everyday voices should be used to synthesize training data.
    \item \textit{Open-source:} Like most researchers, we have a limited budget and want to perform research that is easy to reproduce, so we avoid commercial services like Google's Cloud TTS.
\end{enumerate}
   
For the experiments in this paper, the TTS that best met these criteria was Facebook's VoiceLoop \cite{taigman2017voice}. We used the pre-trained US English model included with the VoiceLoop repo, which has 22 synthetic speakers trained using the VCTK dataset \cite{veaux2017cstr}. We have listened to some of the synthesized audios selected at random and found the VoiceLoop speech to sound fairly natural. However, the synthesized speech does have some flaws: it contains audible vocoder artifacts, punctuation is ignored, and in some instances the model did not correctly pronounce the input text.  Despite these imperfections, the synthesized speech works quite well for training, as we will show.

\section{Experiments}
\label{sec:experiments}

To test our method, we run a number of experiments on two open-source SLU datasets.

\subsection{Datasets}
For the main set of experiments, we use the
Fluent Speech Commands dataset \cite{Lugosch2019}. Fluent Speech Commands is a dataset of 30,043 English audios with 77 speakers, each labeled with ``action'', ``object'', and ``location'' slots. There are 248 distinct sentences, each spoken by multiple speakers in both the training set and validation/test sets.

We also use the Snips SLU Dataset \cite{saade2018spoken}, more specifically the ``smart lights'' near-field subset of the dataset. This dataset is smaller and more challenging than Fluent Speech Commands: it contains numbers and has only 1,660 audios, each corresponding to a different sentence,  so the model is tested entirely on sentences it has never heard before and must generalize to them to achieve high accuracy. Also, the number of slots varies across sentences: for example, the sentence ``Could you turn the lights on please?'' has the label \texttt{\{"intent": "SwitchLightOn", "slots": []\}} with no slots, but the sentence ``Turn the flat light to twelve'' has the label \texttt{\{"intent": "SetLightBrightness", "slots": [\{"entity": "house\_room\_unique", "slot\_name": "room", "text": "flat"\}, \{"entity": snips/number", "slot\_name": "brightness", "text": "twelve"\}]\}} with two slots. 
The dataset is intended to be split into five folds for cross-validation and has multiple speakers, but the splits and speaker identities are not included in the dataset.

\subsection{Models}

\begin{figure}[t]
    \centering
    \includegraphics[scale=0.5,trim={00 0.75cm 13.5cm 1cm}, clip]{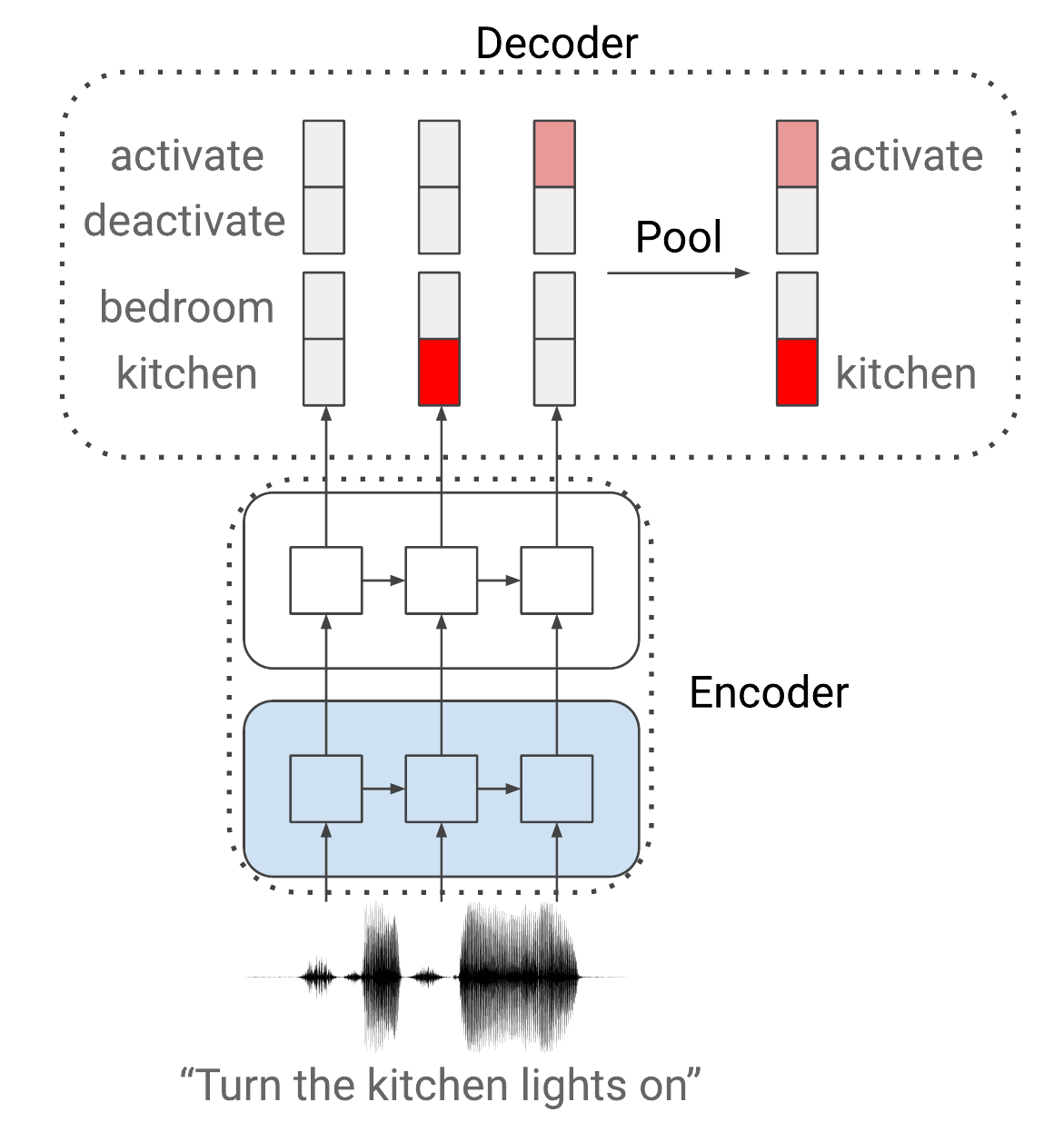}
    \caption{Model with max-pooling decoder. The portion of the model shaded in blue is pre-trained using an ASR task.}
    \label{fig:max-pooling-decoder}
\end{figure}

We use encoder-decoder models in our experiments. The encoder is a deep neural network with multiple convolutional layers and recurrent layers, with max-pooling in some layers to reduce the sequence length. The encoder is pre-trained using the LibriSpeech ASR dataset \cite{librispeech}; more details on how the pre-training is done are given in \cite{Lugosch2019}. The decoder for Fluent Speech Commands is a linear classifier applied to the output of the encoder at each timestep separately, followed by global max-pooling to convert the variable-length sequence of vectors of slot scores into a single vector (Fig. \ref{fig:max-pooling-decoder}). For simplicity, we use the same hyperparameters and transfer learning methodology as were used in the best performing model in \cite{Lugosch2019} across all experiments.

\begin{figure}[t]
    \centering
    \includegraphics[scale=0.5,trim={013.5cm 0cm 0cm 0cm}, clip]{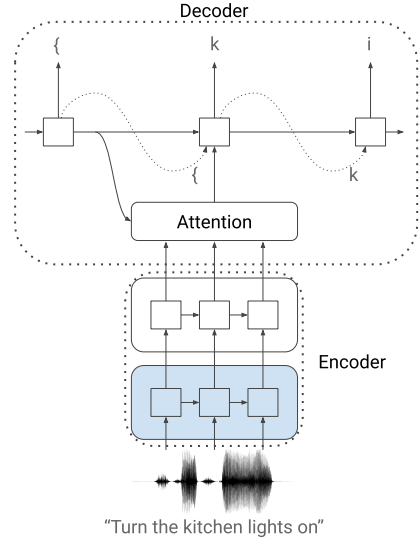}
    \caption{Model with autoregressive decoder used for the Snips SLU Dataset.}
    \label{fig:autoregressive-decoder}
\end{figure}

For the Snips SLU Dataset, since the number of slots varies across utterances,  it is not possible to use the simple max-pooling decoder with a fixed-length output. Instead, we use an attention-based autoregressive decoder \cite{bahdanau2014neural}, as was proposed for SLU in \cite{Haghani2018} (Fig. \ref{fig:autoregressive-decoder}). The decoder uses two gated recurrent unit (GRU) layers of 256 hidden units each \cite{gru}, with key-value attention \cite{transformers}, and sequentially predicts the semantic label string, character by character, using a beam search. We trained autoregressive models on Fluent Speech Commands and used the test accuracy to determine the hyperparameters used in the models for the Snips SLU Dataset. 

\subsection{Results for purely synthetic training sets}\label{ssec:purely_synthetic}


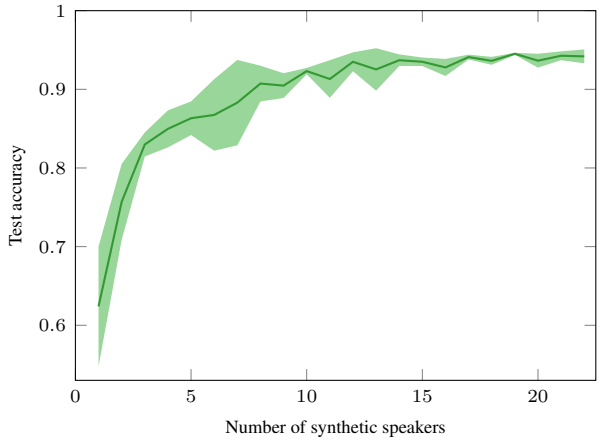
\begin{figure}[t]
    \centering
\hspace*{-0.5cm} 
    \begin{tikzpicture}[font=\fontsize{7}{7}\selectfont]
\begin{axis}[
    ytick={0.6,0.7,0.8,0.9,1.0},
        height=6.5cm,
        width=8.5cm,
        xmin=0,
        xmax=22.5,
        ymin=0.53,
        ymax=1,
        every axis plot/.append style={thick},
        xlabel=Number of synthetic speakers,
        ylabel=Test accuracy,
        x label style={at={(axis description cs:0.5,0.025)},anchor=north},
        y label style={at={(axis description cs:0.1,.5)},anchor=south}
        ]
\addplot[color=black!50!green!80] table[x=x,y=y] {synthetic_only.dat};

\addplot [name path=upper,draw=none] table[x=x,y expr=\thisrow{y}+\thisrow{err}] {synthetic_only.dat};
\addplot [name path=lower,draw=none] table[x=x,y expr=\thisrow{y}-\thisrow{err}] {synthetic_only.dat};
\addplot [fill=black!40!green!40] fill between[of=upper and lower];
\end{axis}
\end{tikzpicture}
    \caption{Test accuracy on Fluent Speech Commands as a function of the number of synthetic speakers.}
    \label{fig:synth_only}
\end{figure}

We first present results for models trained using only synthetic speakers. We used all 22 synthetic VoiceLoop speakers to synthesize all sentences in Fluent Speech Commands\footnote{The synthesized dataset can be downloaded here: \texttt{https://zenodo.org/record/3509828}}. To quantify how many speakers are needed to achieve good accuracy, we train models using the data from one speaker, two speakers, and so on, and report the resulting accuracy. The accuracy is measured on the test set of real speakers in Fluent Speech Commands; if any slot is incorrectly predicted, the entire utterance is deemed incorrect. 

Not every speaker is equally high-quality or useful for training, so the randomly chosen subset of speakers can have a big impact on test accuracy, in addition to other sources of stochasticity, like the initial model weights and the order in which training examples are presented. To reduce the variance of the results, we run each experiment five times using different random seeds, and record the mean and standard deviation of the results.

Fig. \ref{fig:synth_only} shows the test accuracy as a function of the number of speakers. The accuracy increases sharply up to about 15 speakers, and plateaus afterwards, with a very slight upward trend. The conclusion we draw is that one should use all available synthetic speakers if possible, but if synthesis is expensive, or if the resulting dataset is too large to train on exhaustively, it may make sense to incrementally add new synthetic speakers and stop when adding more speakers is not very helpful. In subsequent experiments when using synthetic speakers, we use all 22 available synthetic speakers.

\subsection{Results combining real and synthetic speech}\label{ssec:augment}
We next present results for when the model is trained using real speech and augmented with synthetic speech. We simulate the scenario where only a few real speakers are available by selecting a random subset of speakers from the full training set. The experiments here take longer to run since there are more speakers, so we run each experiment just three times instead of five times. 


\begin{figure}[t]
    \centering
\hspace*{-0.5cm} 
    \begin{tikzpicture}[font=\fontsize{7}{7}\selectfont]
\begin{axis}[
    xtick={0,10,20,30,40,50,60,70},
    ytick={0.94,0.95,0.96,0.97,0.98,0.99},
    legend style={at={(0.72,0.35)},anchor=north},
        height=6.5cm,
        width=8.5cm,
        xmin=0,
        xmax=77,
        ymin=0.93,
        ymax=0.995,
        every axis plot/.append style={thick},
        xlabel=Number of real speakers,
        ylabel=Test accuracy,
        x label style={at={(axis description cs:0.5,0.025)},anchor=north},
        y label style={at={(axis description cs:0.075,.5)},anchor=south}
        ]

\addplot[color=black, dashed] table[x=x,y=y] {vary_real_all_real.dat};
\addplot [name path=upper,draw=none] table[x=x,y expr=\thisrow{y}+\thisrow{err}] {vary_real_all_real.dat};
\addplot [name path=lower,draw=none] table[x=x,y expr=\thisrow{y}-\thisrow{err}] {vary_real_all_real.dat};
\addplot [fill=black!40] fill between[of=upper and lower];

\addplot[color=black!50!green!80, dashed] table[x=x,y=y] {vary_real_all_synth.dat};
\addplot [name path=upper,draw=none] table[x=x,y expr=\thisrow{y}+\thisrow{err}] {vary_real_all_synth.dat};
\addplot [name path=lower,draw=none] table[x=x,y expr=\thisrow{y}-\thisrow{err}] {vary_real_all_synth.dat};
\addplot [fill=black!40!green!40] fill between[of=upper and lower];

\addplot[color=blue] table[x=x,y=y] {vary_real_use_synth.dat};
\addplot [name path=upper,draw=none] table[x=x,y expr=\thisrow{y}+\thisrow{err}] {vary_real_use_synth.dat};
\addplot [name path=lower,draw=none] table[x=x,y expr=\thisrow{y}-\thisrow{err}] {vary_real_use_synth.dat};
\addplot [fill=blue, fill opacity=0.5] fill between[of=upper and lower];

\addplot[color=red] table[x=x,y=y] {vary_real_no_synth.dat};
\addplot [name path=upper,draw=none] table[x=x,y expr=\thisrow{y}+\thisrow{err}] {vary_real_no_synth.dat};
\addplot [name path=lower,draw=none] table[x=x,y expr=\thisrow{y}-\thisrow{err}] {vary_real_no_synth.dat};
\addplot [fill=red, fill opacity=0.5] fill between[of=upper and lower];
\legend{,,,all 77 real speakers,,,,all 22 synthetic speakers,,,,+ all synthetic speakers,,,,+ no synthetic speakers};

\end{axis}
\end{tikzpicture}
    \caption{Test accuracy on Fluent Speech Commands as a function of the number of real speakers.}
\label{fig:vary_num_real}
\end{figure}
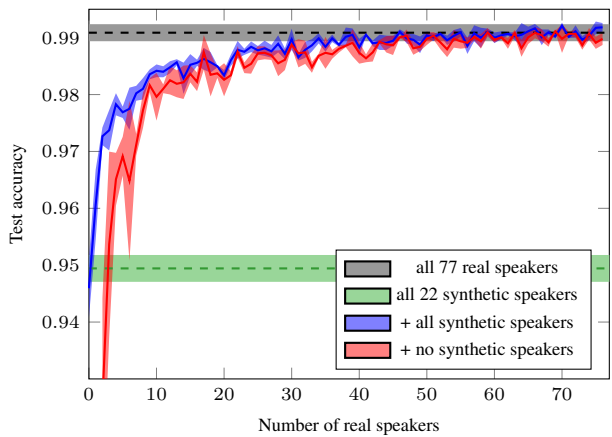

Fig. \ref{fig:vary_num_real} shows the results, presented alongside the accuracy when all 22 synthetic speakers are used (green bar on bottom) and the accuracy when all 77 real speakers are used (grey bar on top). Unsurprisingly, real speech is more useful than synthesized speech. The model trained using only real speech is about 4\% more accurate than the model trained using only synthetic speech (99.1\% $\pm$ 0.1\% versus 94.9\% $\pm$ 0.2\%). Also, with only three real speakers and no synthetic speakers, the model already performs better than when using all 22 synthetic speakers. 

Up to about 40 real speakers, there is unambiguous improvement from including the synthetic speakers in the training set. When using more than 40 real speakers, it is less clear from our experiments if including synthetic speakers is helpful. We measured the difference in accuracy across the number of real speakers with more than 40 real speakers; the accuracy was 0.07\% higher on average when synthetic speakers were included. The difference is not significant, but it at least suggests that it is not harmful to include synthetic speakers even when a large number of real speakers is available.

Finally, we present results for the smaller, more challenging Snips SLU Dataset. Again, we synthesize each sentence using all 22 speakers, which boosts the size of the training set for each fold from 1,328 audios to 30,544 audios. The autoregressive model used for this dataset requires many more steps of stochastic gradient descent (SGD) to fit a dataset than the simpler max-pooling model, so we upsample the real-only dataset so that an equivalent number of SGD steps are taken each epoch for that dataset as for the dataset with synthetic speakers. The model is able to overfit the dataset without synthetic speakers; we therefore record the best test accuracy achieved over the course of training for each fold, instead of the final test accuracy. We also record the best loss, i.e. the average negative log likelihood of the correct semantic label sequence when teacher forcing is used in the decoder. Table \ref{tab:snips} reports these results: both the best accuracy and best loss are significantly better when synthetic speakers are included.

\begin{table}
  \caption{Cross-validation results for Snips SLU Dataset.}
  \label{tab:snips}
  \centering
  \begin{tabular}{l r r r}
    \toprule
    \textbf{Data type} & \textbf{Best accuracy} & \textbf{Best loss}\\
    \midrule
    Real & 65.5\% $\pm$ 2.9\% & 2.81 $\pm$ 0.42\\
    Real + synthetic  & 71.4\% $\pm$ 1.4\% & 1.67 $\pm$ 0.16\\
    \bottomrule
  \end{tabular}
  
\end{table}

\section{Conclusion}
\label{sec:refs}

In this paper, we have shown that it is possible to use synthetic speech to train an end-to-end SLU model. Including synthesized speech in the training set improves accuracy across a variety of settings, in some cases by a large amount. Our results strongly suggest that practitioners should try our method to augment their datasets.

There is still a gap between the performance of a model trained solely on synthetic speech and a model trained on a comparable amount of real speech. In the future, we hope to find ways to reduce this gap, which might include trying other forms of TTS, (automatically) removing badly synthesized utterances from the training set, and combining synthesized speech with other traditional methods for data augmentation, like additive noise and speed perturbation.


\vspace{0.5cm}
\noindent\textbf{Acknowledgments}

Thanks to \mbox{Santi} \mbox{Pascual} and \mbox{Kyle} \mbox{Kastner} for helpful discussions with us about multi-speaker TTS, and thanks to \mbox{Alice} \mbox{Coucke} for help with using the Snips SLU Dataset.


\bibliographystyle{IEEEbib}
\bibliography{strings,refs}

\end{document}